# A long-term investigation on the effects of (personalized) gamification on course participation in a gym

Maximilian Altmeyer[a], Marc Schubhan[a], Antonio Krüger[a] and Pascal Lessel[a]

[a]*German Research Center for Artificial Intelligence (DFKI), Saarland Informatics Campus, Saarbrücken, Germany*

**Abstract**
Gamification is frequently used to motivate people getting more physically active. However, most systems follow a one-size-fits-all gamification approach, although past research has shown that interpersonal differences exist in the perception of gamification elements. Also, most studies investigating the effects of gamification are rather short, although it has been shown that gamification can suffer from novelty effects. In this paper, we address both these issues by investigating whether gamification elements, integrated into a fitness course booking system, have an effect on how frequently users participate in fitness courses in a gym (N=52) over a duration of 275 days (548 days including baseline). Also, the gamification elements that we implemented are tailored to specific Hexad user types, which allows us to investigate whether using suitable gamification elements leads to an increased course participation. Our results show that gamification increased the participation in fitness courses significantly and that users who received a suitable set of gamification elements–according to their Hexad user type–increased their participation significantly more than others.

## 1. Introduction

One of the main health concerns in today's society is the lack of physical activity, since an increasing number of people are leading sedentary lifestyles [1]. In fact, more than 60% of Americans are not physically active regularly and 25% are completely inactive, according to a US governmental report [2]. Consequently, this leads to a broad range of health issues, such as cardiovascular diseases, obesity and various other chronic illnesses [3].

Therefore, encouraging people to lead an active lifestyle and to increase their physical activity level has a great potential for private and public health [4]. As a result, systems and interventions aiming at encouraging physical activity have been designed, implemented and studied [5]. Often, such systems make use of gamification, *the use of game elements in non-game contexts* [6], to motivate and engage users and thus help them reaching their fitness goals [5, 7]. Gamification interventions mostly use a "one-size-fits-all" approach (i.e., using the same set of gamification elements for all users) [8, 9, 10]. However, previous research has demonstrated that there are interpersonal differences in the perception of gamification elements [11], which poses a threat to such static approaches. Therefore, personalization and understanding the factors explaining why gamification works recently emerged as thriving fields in gamification research [12].

As a result, Marczewski [13] established the Hexad user types model, which has been developed specifically to explain user preferences in gamified systems [14, 15]. It describes six user types, based on Self-Determination Theory [16]. Tondello et al. [11] developed a questionnaire to assess Hexad user types and subsequently demonstrated its validity [17]. Although the Hexad model is relatively new to the field, it has received much attention and was used in various domains, including physical activity [18], education [19], energy conservation [20] and health [14]. These studies consistently found that the *perception* of gamification elements is correlated to Hexad user types. The Hexad model was also empirically shown to be advantageous in explaining interpersonal differences in the perception of gamification elements [21]. However, previous research did not allow participants to interact with gamified systems but instead mostly let them rate their perception of gamification elements based on e.g. textual descriptions or storyboards [22, 23]. Consequently, it remained unclear whether personalization has actual effects on the behavior of users in an in-the-wild setting, when giving them the chance to interact with gameful applications. Besides the need for personalization, gamification research lacks empirical evidence for its long-term success [5, 24]. Based on the results of Aldenaini et al. [24], we consider studies as longitudinal when being longer than one year. The authors found that a huge majority of previous studies had a rather short duration [24]. Thus, the question whether gamification has a lasting impact on the physical activity levels of participants remains unclear. This poses the question if gamification is a suitable approach to motivate users to change their behavior sustainably.

---

In this paper, we contribute to both these aspects–the effects of personalization based on Hexad user types in-the-wild as well as the long-term effects of gamification. We integrated gamification elements, which are personalized to a specific set of Hexad user types, into an existing course booking system of a local gym and asked users to voluntarily fill out a questionnaire to assess their Hexad user type. After two years (one year baseline and one year using gamification), we received a completely anonymized dataset containing information about the interaction and booking behavior of 52 users over the past two years. Our results show that users booked significantly more fitness courses during the year in which gamification was active compared to the baseline. Also, we found that users who received a suitable set of gamification elements according to their Hexad user type, booked significantly more fitness courses than those who did not receive a suitable set of gamification elements. Thus, our contribution is two-fold. First, we provide empirical evidence for the long-term success of gamification in the physical activity domain. Second, we demonstrate the practical impact of personalization based on the Hexad user types model.

## 2. Related work

We start by presenting past research that utilized gamification elements to encourage users to become more physically active. Next, we present the state-of-the-art in personalized gamification, with an emphasis on the Hexad model. We conclude the section by summarizing the main findings and framing our contribution.

### 2.1. Gamification and physical activity

The fact that gamification can increase the motivation to perform physical activity has been shown by numerous investigations in the past. For instance, UbiFit Garden [25], which shows a virtual garden on the wallpaper of users' mobile phones, was shown to increase their activity levels in a 3-months experiment. The state of this virtual garden is connected to physical activity by conveying progress towards activity goals through flowers and butterflies growing and appearing. Building upon social comparison, StepStream [26] additionally uses goals based on the performance of other users to enhance their motivation to become physically active. The system shows a social stream on a website, in which achievements are visualized, when students reach their daily step goal. Weekly sessions were held at school, where students were encouraged to use this website. Although the system did not lead to an increase in step counts in a 4-weeks field trial, attitudes towards physical activity became more positive. Similar to our setting, Altmeyer et al. [27] investigated the effect of showing step counts of users on a public display, which was located in a gym, in addition to a mobile application. In a 4-weeks user study, they found that the public display and the increased social relatedness were the cause for a significant increase of step counts. Chen et al. [28] implemented a gamified system called "HealthyTogether" in which users were paired and experienced different social settings including a competitive, collaborative and a hybrid setting. They found that all settings increased the activity of users and that collaborative gamification elements led to stronger improvements than competitive ones. Fish'n'Steps [29] links users' step counts to the growth and emotional state of a virtual fish to motivate users to walk more. In a 14-weeks user study in an office environment, all participants were able to see their personal fish tank, while half of the participants additionally were grouped in teams, which had their own fish tanks. These team fish tanks were shown on a public display, thus introducing social comparison and collaboration. Although the system was well-received, the study found no differences in the amount of steps walked between these two conditions.

Recent literature reviews by Aldenaini et al. [24] as well as Koivisto and Hamari [5] support the fact that the success of gamified interventions differs. Aldenaini et al. [24] reviewed 170 papers aiming to change the physical activity behavior of users. While they found that 51% of studies were fully successful, 49% were only partially successful or even unsuccessful, which strengthens the need to investigate whether personalization might help to increase the success of gamified systems. They also found that most studies had a rather short duration, with roughly 40% of studies being shorter than one month and 70% being shorter than three months. Thus, the authors conclude that the long-term effects of systems encouraging physical activity need further research. In line with this, Koivisto and Hamari systematically reviewed 16 comparison studies on gamification of physical activity [5]. They found that mostly positive outcomes of gamification were reported but remark that more rigorous study designs would lead to less optimistic findings. Also, the authors found that most studies relied on self-reported data instead of objective measurements and that many studies might be affected by novelty effects due to their short evaluation phase. In contrast to the aforementioned studies, we analyze data from two years instead of relying on rather short time frames as well as use objective measurements in contrast to self-reported data.

### 2.2. Personalized gamification

To understand why gamification is successful or not, researchers investigated several factors which might play a role in the perception and effectiveness of gamifica-tion elements. Jia et al. [9] investigated the influence of personality traits and found that they explain the perception of certain gamification elements, e.g. the authors found that "extroversion" positively impacts the perception of points and levels. Similarly, Orji et al. [30] investigated whether personality traits explain differences in the the perceived persuasiveness of persuasive strategies. They created storyboards explaining each strategy in the context of unhealthy alcohol behavior and found similar effects as Jia et al. [9]. Besides personality, researchers have considered age as a potential factor for personalization. As such, Birk et al. [31] found that play motives and preferences changes in old age, i.e. that with increasing age, participants focus more on enjoyment instead of performance. Similarly, Kappen et al. [32] found that personalizing gameful applications to support physical activity is important, because age-specific challenges need to be accounted for. In a follow-up work, Kappen et al. [33] conducted an eight-week study about how gamification influences older adults in the context of physical activity. They found that gamification elements should be customized to the older adult's needs and conclude the paper with design guidelines. In addition to age, gender-wise differences have been demonstrated by Oyibo et al. [34], who found that competition and virtual rewards are preferred by male participants. Besides demographic factors and personality traits, gamification research suffered from missing theoretical models which are specifically developed for the purpose of personalized gamification until the Hexad user type model was introduced by Marczewski [13]. It consists of the following six user types, based on Self-Determination Theory [16]:

**Philanthropists ("PH")** are socially-minded, like to take responsibility, and share their knowledge with other users. Their main motivation is *purpose*.
**Socialisers ("SO")** are also socially-minded but are more interested in interacting with other users. Therefore, they are mainly driven by *relatedness*.
**Free Spirits ("FS")** like to explore and act without external control, with *autonomy* being most important for them.
**Achievers ("AC")** enjoy overcoming challenging obstacles and mastering difficult tasks. They are motivated by *competence*.
**Players ("PL")** are focused on their own benefits, and are driven by the will to win and earn external rewards. Hence, *extrinsic rewards* are most important for them.
**Disruptors ("DI")** enjoy testing a system's boundaries and are motivated by triggering *change*, either positive or negative.

As a foundation for further research, Tondello et al. [11] developed a questionnaire to assess Hexad user types, which has been validated recently [17]. In addition, the usefulness of the Hexad model in explaining user preferences in gamified systems was demonstrated by numerous interventions across various domains. In the health domain, Orji et al. [14] found that a users' Hexad type explained the perceived persuasiveness of several persuasive strategies. In an educational context, Mora et al. [19] investigated the potential of using the Hexad model to personalize learning experiences and found that the approach that utilized the Hexad model to personalize the game design elements yielded higher engagement of the students. In the context of energy efficiency at the workplace, Kotsopoulos [20] similarly found that the perception of gamification elements is explained by Hexad user types. Most relevant to our paper, the Hexad model and potential correlations between user types and the perception of gamification elements were investigated in the fitness context by Altmeyer et al. [18]. The authors demonstrated the validity of previously found correlations, as their results showed that the perception of gamification elements is correlated to similar Hexad user types. Providing further support for the Hexad model, Hallifax et al. [21] compared the Hexad user types model to other models (including the BrainHex [35] model, and the Big-5 personality traits model [36]) and found that it is the most suitable typology for this purpose, since most of the correlations to gamification elements that were found by the authors are in line with the definitions of the Hexad user types.

### 2.3. Summary

To sum up, previous research has demonstrated that gamification has great potential to help people achieving their fitness goals and enhancing their physical activity. However, a major issue in gamification research in the domain of physical activity was shown to be the short duration of user studies, which poses the question if gamification was successful because of the motivational pull of gamification elements or because of novelty effects. We contribute to this open question by analyzing data of two years (one year baseline and one year intervention).

In addition, it was also shown that interpersonal differences exist in the perception of gamification elements, which might explain why the success of gamification differs substantially across the reported interventions. To account for these interpersonal differences, personalization was shown to be important for gameful systems. The Hexad user model is the only typology to date which aims to explain user preferences in gameful systems, which is why we decided to use it to investigate personalization-related research questions. While previous studies have relied on subjective measures and self-reports of user preferences based on storyboards or textual descriptions, we investigate the effects on actual user behavior in-the-wild and thus contribute empirical findings regarding the usefulness of the Hexad model in personalizing gamified applications. Moreover, due to the long study duration, we are able to shed light on the usefulness of personalized gamification in the long-run, which has not been investigated before, as far as we know.

## 3. Fitness course booking system and gamification elements

To investigate the effectiveness of gamification elements and personalization on the amount of booked fitness courses, we were allowed to integrate gamification elements into an existing web-based course booking system, which will be described in the following.

### 3.1. Gym and course booking system

The local gym in Germany, which we collaborated with, does solely offer courses, i.e. users have to register beforehand for a course manually and are not allowed to participate in any fitness activity when not registered. Thus, course bookings are binding. To book fitness courses, users have to login to the website of the gym with personal credentials. Next, they can see which courses are offered in the following two weeks and can freely decide which courses to book. It is important to note that courses need either to be payed manually or users may have a subscription. In case of the latter, users pay a monthly fee and are allowed to participate in a fixed number of courses per week. If they book more courses in a week than is covered by their subscription, the respective courses will be charged manually. This means that each booked course needs to be payed.

There are different types of courses. While some focus on strength, others focus on endurance or combine fitness activities and exercises with teaching the proper technique. The number of people in a course typically is bound to ten or twelve. Also, courses are offered throughout the day and usually take an hour. The earliest courses start at around 7am and the latest courses end at 9pm. There are courses on each day of the week, but the number of courses per day is slightly reduced on Sundays.

### 3.2. Gamification elements

We included four different gamification elements, which are described in the following. Additionally, we implemented an option to upload profile pictures and set a username, which would be shown on the leaderboard and when booking a course (such that users could see which other users take part in the course). This was done to allow for social comparison outside of the leaderboard. The selection of gamification elements orientated on frequently used gamification elements in fitness contexts [5].

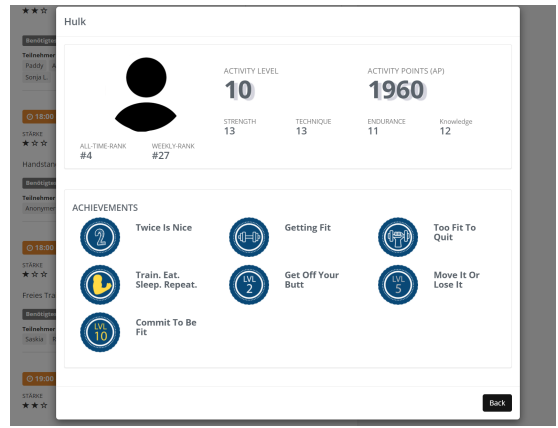

**Figure 1:** Profile information, after clicking on a user when booking a course or on the leaderboard

We also implemented an option to stay anonymous. This means that the username is not shown when booking courses and is also hidden in the leaderboard. Per default, users are anonymous and need to enable that they would like to be shown on the leaderboard/when booking courses in their profile settings.

#### 3.2.1. Activity points

Points have been shown to positively affect **Players** [11, 13] and **Socialisers** [14, 18]. We implemented a point system based on so-called Activity Points ("AP"). When users book a course, they receive 10 AP. The current amount of earned AP is permanently shown on the left side of the navigation section as well as when users navigate to their profile page (see Figure 2). When collecting APs, users make progress towards reaching their next activity level, which is explained in the following section.

#### 3.2.2. Levels

Levels and progression have been shown to be particularly motivating for **Achievers** and **Players** [11, 13]. We introduced the concept of activity levels, meaning that collecting Activity Points leads to increases in the activity level of a user. The activity level is visualized together with the current amount of Activity Points in a users' profile page as well as permanently on the navigation side bar. Below the current activity level, a progress bar indicates the current progress towards the next activity level and shows how many Activity Points are missing. The number of Activity Points to reach the next activity level grows logarithmically with increasing levels.

Besides the activity level, we also introduced four different attribute levels: The Strength, Endurance, Technique and the Knowledge levels. In consultation with the professional fitness trainers, all courses were rated in terms of these attributes, such that users who book a certain course, receive attribute points and make progress in corresponding attribute levels (depending on the type of course). When booking a fitness course, we added this information such that users know how to improve their attributes and to give them an indication whether their training reflects their personal fitness goals. Besides showing the fitness attributes, we visualized the users who have booked the course (as stated before). When clicking on a user, their profile picture, current rank on the leaderboard, unlocked badges, activity level and the level in each attribute is shown (if the user is not anonymous), see Figure 1.

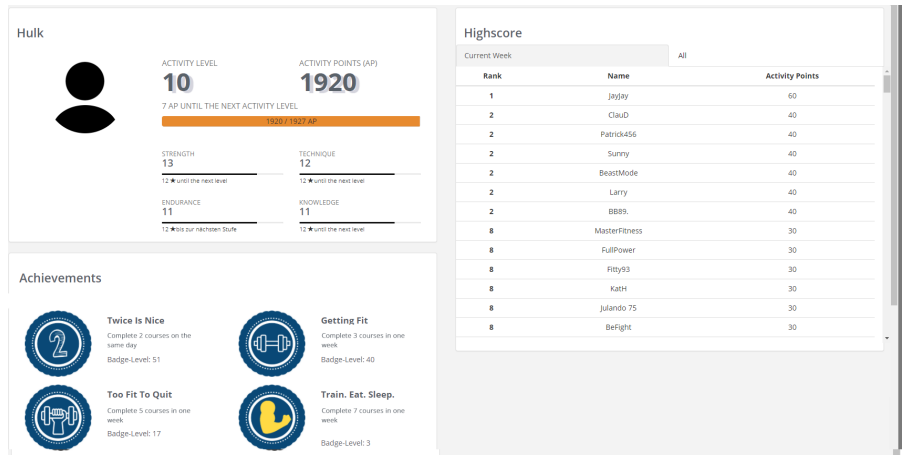

**Figure 2:** Gamification elements shown in the personal profile of the user. The user interface shows the current activity level and the level for each attribute as well as the current progress towards the next level. Also, Activity Points, unlocked badges and the leaderboard are shown. If users have the same score, they are ranked on the same position.

### 3.2.3. Badges

This gameful design element is especially suitable for **Achievers** as it builds on the concept of mastery [13]. Previous research has shown that the perception of Badges is positively correlated to the **Achiever** and **Player** user types [11]. We integrated nine different badges, of which three are triggered by making progress in the activity level, four by completing a certain number of courses per week and two by participating in particularly early or late courses. Figure 2 shows a subset of badges offered.

### 3.2.4. Social competition / leaderboard

**Players** and **Socialisers** were shown to be particularly driven by Social Competition and Leaderboards [11, 13]. We implemented a weekly and an all-time leaderboard to both allow new users to keep up with others and also reward long-term participation. The leaderboard was shown on the profile page (see Figure 2) and the current rank of users was also shown when clicking on a certain user when booking a course (see Figure 1). When clicking on an entry in the leaderboard, the same dialog opened as when clicking on a user when booking a course.

## 4. Evaluation

To investigate the long-term effects of gamification on the number of booked courses as well as the role of personalization based on Hexad user types, we analyzed a fully anonymous dataset after the gamification elements were active for one year (*Gamification* phase) on the booking system and compared it to the year before gamification was introduced (*Baseline* phase).

### 4.1. Method

After the gamification elements were activated, users of the booking system were asked to fill out the validated Hexad user types questionnaire [17] voluntarily. They also had the option to skip filling out the questionnaire. When selecting users to be analyzed, we had the following criteria: First, users should be registered for at least two years, i.e. before 2018-10-08 (we considered data between 2018-10-08 and 2019-10-07 as Baseline). Second, users should have booked at least one course in the month before the end of the Baseline and before the end of the Gamification phase (the Gamification phase started on 2019-10-08 and ended on 2020-10-07). This was done to decrease the chance of including users who quit going to the gym during the study duration. Lastly, we only included users who voluntarily filled out the Hexad user types questionnaire completely. To investigate the effects of personalization, we split participants into two groups. Since the gamification elements described in Section 3.2 were shown to be particularly suitable for Achievers, Players and Socialisers [11, 13], users scoring highest on at least one of these traits were matched to a group of users who received a suitable set of gamification elements. This group was compared to the remaining users, for who the implemented gamification elements should not be particularly suitable, according to the Hexad model. Data was analyzed using paired/unpaired t-tests or the non-parametric counterpart, when assumptions were not met (determined by conducting Levene's/Shapiro-Wilk tests).

### 4.2. Hypotheses

We had the following hypotheses:

**H1:** The number of booked courses per participant is significantly higher in Gamification than in Baseline.

**H2:** Users scoring highest on the Hexad user types Achiever, Socializer or Player–and thus receive a suitable set of gamification elements–increase their number of booked courses significantly more than other users.

**H1** is based on previous research demonstrating that gamification leads to an increased physical activity [27, 37, 38]. Thus, we expect to find similar effects. In contrast to previous work, the study duration is much longer in our study. Therefore, we expect to find a smaller effect than was reported in previous work, because novelty effects decrease over time. Since previous research conducting rather short studies has found medium to large effect sizes [27], we calculated an a-priori power analysis to detect a small to medium sized effect of $d_z=.40$ with a power of 80%, thus revealing a minimum number of 41 participants. **H2** is based on findings of past research, showing that there are correlations between Hexad user types and the perception of gamification elements [11, 18]. Thus, we expect that these self-reported preferences should be reflected in the behavior of users, i.e. that receiving gamification elements that are suitable to the highest Hexad user types should have an effect on the behavior. Given the minimum number of 41 participants as calculated by the aforementioned power analysis, we are able to find large effects of $d=.80$ with a probability of 80%. **H1** is evaluated using paired tests while **H2** is evaluated using unpaired tests.

### 4.3. Dataset

We received the aggregated number of booked courses for each month and per study phase for each eligible user together with the information whether users scored highest on the Hexad factors Achiever, Player or Socialiser. Besides that, we received the following aggregated information: the average scores of all Hexad factors, the average levels, the aggregated number of users who decided to provide a username and the average number of unlocked badges. We did not receive any non-aggregated data nor personal information such as age or gender such that the dataset can be considered fully anonymous. This was important to prevent any GDPR related issues. Due to the COVID-19 pandemic and the resulting lockdown in Germany, we had to exclude data between 2020-03-02 and 2020-06-01. To ensure the comparability to the Baseline phase and prevent seasonal effects, we excluded data from the same timespan in the Baseline, i.e. between 2019-03-02 and 2019-06-01.

### 4.4. Results

Overall, 52 eligible users were considered. The Hexad user types average scores are similar to the ones reported in the validation study of the Hexad questionnaire by Tondello et al. [17]. Philanthropists showed the highest average scores (M=23.79, SD=3.15), followed by Achievers (M=23.60, SD=3.33), Socialisers (M=22.75, SD=3.62) and Free-Spirits (M=22.13, SD=3.55). Players (M=19.60, SD=5.34) and Disruptors (M=14.37, SD=4.87) followed with lower average scores. Based on the Hexad user type scores, our sample consisted of 33 users who received a suitable set of gamification elements (i.e. scoring highest on Achiever, Player or Socialiser) and 19 users who did not receive a suitable set of gamification elements (i.e. who did not score highest on Achiever, Player or Socialiser). The Achiever, Player and Socialiser scores of users who received suitable gamification elements were on average significantly higher than the respective scores of users who did not receive a suitable set of gamification elements, with large effect sizes of $d>0.5$ (Achiever: $t(25.11)=2.25$, $p<0.05$, $d=0.74$; Player: $t(40.44)=1.88$, $p<0.05$, $d=0.53$, Socialiser: $t(31.86)=2.40$, $p<0.05$, $d=0.73$). Users unlocked 4.60 (SD=1.80) badges and a substantial majority of 92% actively decided to be shown on the leaderboard by selecting a nickname.

#### 4.4.1. Effect of gamification on course bookings

To analyze whether gamification had an effect on the number of booked courses (**H1**), we compared the number of booked courses per day between Baseline and Gamification. In the Baseline phase, users booked 0.28 courses on average per day (Mdn=0.28, SD=0.14). This number significantly increased in the Gamification phase (Z=491.00, p=0.036, d=0.31) to an average of 0.30 courses per day (Mdn=0.29, SD=0.15). Thus, we derive result **R1: The number of booked courses per day is significantly higher in Gamification than in Baseline**.

In addition to comparing the full Baseline versus the full Gamification phase, we also compared the number of booked courses on a monthly basis between Baseline and Gamification phase. This was done to abstract from the fact that the time of the year might have a general influence on the behavior and motivation of users to participate in fitness courses in a gym. When looking at Figure 3, it can be seen that the number of booked courses is descriptively higher in all months (but February, which might be due to COVID-19) in the Gamification phase. Significant effects were found for five out of nine months (October: t(51)=-3.38, p<0.01, d=0.47; December: t(51)=2.13, p<0.05, d=0.29; June: t(51)=1.95, p<0.05, d=0.27; August: t(51)=1.76, p<0.05, d=0.24; September: t(51)=1.68, p<0.05, d=0.23). This is summarized as **R2: In the month-by-month comparison, the number of booked courses per day is significantly higher in Gamification for the majority of months**.

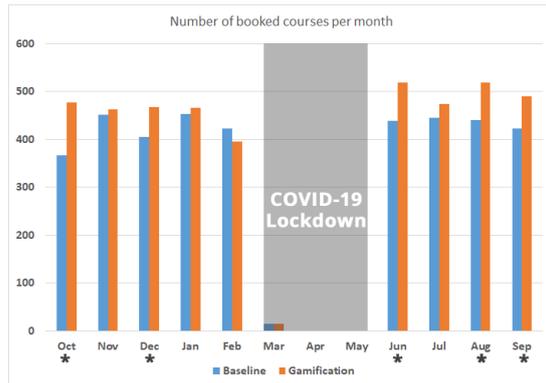

**Figure 3:** Number of booked courses per month. Blue bars (left side) represent the number of bookings in Baseline, orange bars (right) in Gamification. An asterisk represents a significant difference between Gamification and Baseline

#### 4.4.2. Effect of personalization using hexad user types on course bookings

Next, we analyzed if users scoring highest on the Achiever, Socialiser or Player factor of the Hexad –and thus received a suitable set of gamification elements–were more driven by the gamification elements than other users. Therefore, we calculated the ratio between the average number of courses booked per day in Baseline and the average number of courses booked per day in Gamification for each user, to have a relative number indicating the difference in booked courses per day between Baseline and Gamification. This relative metric was chosen to abstract from the fact that users might have different subscriptions, which might introduce a bias to comparing the users who received a suitable set of gamification elements against users who did not. We then split users into a group that received suitable gamification elements (according to previous literature on correlations between Hexad user types and gamification elements, as described in Section 3.2; N=33) and a group that did not receive suitable gamification elements (N=19).

On average, users who did not receive a suitable set of gamification elements have a ratio of 1.01 (SD=0.29). This indicates that the number of booked courses remained almost the same in the Gamification phase. When considering users who received a suitable set of gamification elements, according to their Hexad user type, the average ratio is 1.31 (SD=0.90). This shows that these users increased their number of bookings in Gamification by more than 30% on average. When comparing this ratio between the users who received a suitable set of gamification elements and those who did not, we found a significant difference (t(42.31)=1.78, p<0.05, d=0.41). Thus, we derive **R3: Users who received a suitable set of gamification elements improved their participation during Gamification significantly more than others**. This suggests that personalization had an effect on the number of courses that users booked.

Since users who received a suitable set of gamification elements increased their number of course bookings significantly more than others, we also investigated whether the results that were found regarding the impact of gamification (**R1, R2**) still persist among users who did not receive suitable gamification elements as compared to those who did. Indeed, for users who did not receive a suitable set of gamification elements, none of the significant differences in course bookings reported in Figure 3 were found (i.e. there is no significant difference between Gamification and Baseline for users who did not receive suitable gamification elements), whereas the same significant differences where found for users who received a suitable set of gamification elements. Thus, we establish **R4: The significant differences in the month-by-month comparisons between Gamification and Baseline seem to be attributable to users who received a suitable set of gamification elements**. Furthermore, we analyzed whether there are differences in the amount of interactions with gamification elements. We found that more users who received a suitable set of gamification elements updated their profile (2.09 vs. 1.37 times), and wanted to show their name on leaderboards (94% vs. 89%), as compared to user who did not receive suitable gamification elements, without reaching significance. Regarding the amount of unlocked badges, we found a significant increase (4.94 vs. 4.00) among users receiving suitable gamification elements (t(33.38)=1.79, p=<0.05, d=0.54). These interaction-related results lead to **R5: User who received a suitable set of gamification elements unlocked more badges than others**.

### 4.5. Discussion and limitations

The findings show that users increased their participation significantly during the year in which gamification was activated (**R1**). Also, when analyzing the number of booked courses per day on a monthly basis, it could be seen that the number of booked courses per day was higher in all months but February, with five out of nine months reaching significance (**R2**). This adds further to the fact that gamification affected users positively, apparently even in the long-run. We see **R1** and **R2** as supporting evidence for **H1:The number of booked courses per participant is significantly higher in Gamification than in Baseline**. On a more abstract level, these results contribute novel insights into the long-term effectiveness of gamification, which has been controversially discussed in the field [5, 12, 24].

We found that users who received a suitable set of gamification elements based on their Hexad type increased their participation in fitness courses significantly more than users who did not receive particularly suitable gamification elements (**R3**). In fact, users receiving suitable gamification elements increased their participation by more than 30%, while other users did not increase their participation considerably. Furthermore, when only considering users who did not receive a suitable set of gamification elements, the significant differences in the month-by-month comparisons between Gamification and Baseline disappear (**R4**). This together with **R3** suggests that the increased participation in Gamification (**H1**) might actually be caused by the group of users receiving suitable gamification elements, which undermines **H1** to a certain extent. Consequently, this could mean that the suitability of gamification elements plays a substantial role in the success of gamification. This poses the question, if the variety of positive, neutral or negative outcomes in previous literature [5, 8, 24] is due to the selection of suitable or unsuitable gamification elements. We also found that users for whom the gamification elements were suitable unlocked significantly more badges and interacted (descriptively) more with gamification-related features of the system (**R5**). **R3**–**R5** are important findings, since previous research has not considered behavioral data but solely focused on self-reported preferences, as far as we know. We see these results as supporting evidence for **H2: Users scoring highest on the Hexad user types AC, SO or PL–-and thus receive a suitable set of gamification elements–-increase their number of booked courses significantly more than other users**.

Our study has several limitations which should be considered. First, we would like to acknowledge that the users we considered had to pay for every single course they booked, which limits the autonomy of their decision. Thus, it could be that the effect sizes we reported are different in contexts were users have a free choice of how much physical activity they would like to perform. Also, we selected participants who were participating in at least one course both in the last month of the Baseline as well as in the last month of the intervention phase. While this ensures that users who quit the gym due to external factors (such as changing the place of residence) are not considered in the sample, users who quit due to other reasons are also excluded. Therefore, future work should follow a study design which separates intervention and control groups, instead of doing a within-subjects study or ask users who quit about their reasons. In addition, it should be noted that our target group were users who already decided to visit the gym, which might have an impact on the success of gamification elements as reported in past research [18, 39]. Regarding **H2**, it should be considered that we used a dichotomous approach in deciding whether a certain user received suitable gamification elements or not, which was based on whether the users scored highest on the Achiever, Socialiser or Player factor of the Hexad (because the gamification elements that we implemented were shown to be perceived particularly well among these user types). This has the advantage of an increased statistical power (due to less factors to differentiate), but comes at the cost of potential simplification (since the Hexad consists of six factors) and should be considered when interpreting our findings. Lastly, it should be noted that the Gamification phase was affected by the COVID-19 pandemic. Due to a nation-wide lockdown and the closure of the gym, we had to remove roughly three months from the Gamification phase. To account for this limitation and ensure the month-wise comparability of the data, we removed the corresponding days from the Baseline phase. However, we do not know in how far the pandemic has influenced the behavior of users. The fact that the number of booked courses was (descriptively) lower solely for the month February in the Gamification phase suggests that the COVID-19 pandemic already had an effect on the behavior of users in February 2020. Furthermore, we do not know whether the closure might have led to users booking courses more frequently when the gym was opened again.

## 5. Conclusion and future work

In this paper, we investigated the long-term effects of gamification and whether personalization based on Hexad user types has an influence on the behavior of users in the context of course participation in a gym. We implemented gamification elements, which are particularly suitable for Achievers, Players and Socialisers on an existing course booking website and received a fully anonymized dataset over a period of two years. Our findings show that gamification significantly increased the participation of users in fitness courses and thus seems to have a lasting impact on the behavior of users. Furthermore, we found that users scoring highest on the Achiever, Socialiser or Player factor of the Hexad and thus received suitable gamification elements on the platform (points, leaderboard, badges, levels), increased their participation in fitness courses significantly more than users who did not receive a suitable set of elements. Our findings also suggest that the success of gamification is moderated by the selection of gamification elements and their suitability for the target audience, which provides potential explanations for the varying success of gamification in previous studies.

Future work should investigate the lasting impact of gamification in physical activity contexts in which users are free to decide to what extent they would like to increase their activity and are not limited by financial factors. Also, future work should be conducted on whether personalizing gamified systems to Hexad user types affects the behavior of users by dynamically adapting the gamification elements instead of providing a static set of gamification elements like was done in this study, to get a more holistic picture on the effectiveness of personalization based on Hexad user types.